\documentclass{hotnets2020}

\usepackage{times}  
\usepackage{hyperref}
\usepackage{xcolor}
\usepackage{ifthen}
\usepackage{xspace}

\usepackage{soul} 
\usepackage[all]{nowidow}
\hypersetup{pdfstartview=FitH,pdfpagelayout=SinglePage}
\newcommand{\mypara}[1]{\smallskip\noindent{\bf {#1}.}~}
\newcommand{\showComments}{yes}

\newcommand{\note}[2]{
\ifthenelse{\equal{\showComments}{yes}}{\textcolor{#1}{#2}}{} }

\begin{document}

\newcommand{\sys}{\mbox{NSinC}\xspace}

\newcounter{note}[section]
\newcommand{\fixmecolor}{red}
\newcommand{\notecolor}{blue}
\renewcommand{\thenote}{\thesection.\arabic{note}}
\newcommand{\vyas}[1]{\refstepcounter{note}{\bf \textcolor{\notecolor}{$\ll$Vyas~\thenote: {\sf #1}$\gg$}}}
\newcommand{\dk}[1]{\refstepcounter{note}{\bf \textcolor{red}{$\ll$DK~\thenote: {\sf #1}$\gg$}}}
\newcommand{\alan}[1]{\refstepcounter{note}{\bf \textcolor{red}{$\ll$alan~\thenote: {\sf #1}$\gg$}}}
\newcommand{\ga}[1]{\refstepcounter{note}{\bf \textcolor{\notecolor}{$\ll$George~\thenote: {\sf #1}$\gg$}}}
\pagenumbering{arabic}
\title{Unleashing  In-network Computing on Scientific Workloads}
\author{Daehyeok Kim$^{1}$, Ankush Jain$^{1}$, Zaoxing Liu$^{1}$, George Amvrosiadis$^{1}$, Damian Hazen$^{2}$, \\Bradley Settlemyer$^{3}$, Vyas Sekar$^{1}$\\
{\normalsize $^{1}$Carnegie Mellon University, $^{2}$Lawrence Berkeley National Laboratory, $^{3}$Los Alamos National Laboratory}}

\maketitle

\begin{abstract}


Many recent efforts have shown that in-network computing can benefit various datacenter applications. In this paper, we explore a relatively less-explored domain which we argue can benefit from in-network computing: scientific workloads in high-performance computing. By analyzing canonical examples of HPC applications, we observe unique opportunities and challenges for exploiting in-network computing to accelerate scientific workloads.  In particular, we find that the dynamic and demanding nature of scientific workloads is the major obstacle to the adoption of in-network approaches which are mostly open-loop and lack runtime feedback. In this paper, we present NSinC (Network-accelerated ScIeNtific Computing), an architecture for fully unleashing the potential benefits of in-network computing for scientific workloads by providing closed-loop runtime feedback to in-network acceleration services. We outline key challenges in realizing this vision and a preliminary design to  enable  acceleration  for scientific applications.

\end{abstract}

\section{Introduction}
\label{sec:intro}




Recent advances in programmable hardware switches~\cite{www-tofino, www-xpliant} and NICs~\cite{Mellanox17, www-netronome} have enabled  the network data plane to move beyond its traditional role of packet forwarding. Today, we have   more sophisticated  capabilities to process packets and offer advanced capabilities to accelerate both network- and application-level functions. 
This emerging trend toward \textit{in-network computing} enables new opportunities to improve performance and lower operational costs of data center infrastructure~\cite{hotnets17-daiet}. 
Indeed, recent efforts have shown that  datacenter applications, such as key-value stores~\cite{sosp17-netcache, sosp17-kvdirect}, machine learning~\cite{tr-switchml}, and network functions~\cite{sigcomm17-silkroad, sigcomm18-sonata} can benefit from in-network computing, and a recent paper~\cite{hotos19-innet} has provided guidelines for judiciously using in-network computing.

In this paper, we focus on a relatively less  explored domain that  we argue stands to benefit from in-network computing: scientific workloads executing on high-performance computing (HPC) platforms. Some canonical examples of such workloads  include simulation tools  for designing new materials~\cite{parsplice}, genome sequencing~\cite{hipmer}, and simulating magnetic reconnections~\cite{vpic13, vpic15}.  At a high level, these 
 workloads  simulate various physical phenomena that are difficult or impossible to physically observe, as we show in \S\ref{sec:canonapps}. 
 We highlight  novel and as-yet-unrealized opportunities for leveraging programmable network devices to significantly accelerate these workloads. As the workloads perform frequent communications between nodes for data movement and synchronization, we find that there are many opportunities to accelerate them using information directly collected from inside the network for better load-balancing, data partitioning, and domain-aware caching  (\S\ref{sec:opportunities}). 

While using in-network computing is   appealing, we observe that simply adopting existing approaches designed for other application domains (e.g., datacenter type applications)   fail to  fully realize these potential benefits.  The key reason, as we discuss in~\S\ref{sec:case}, is that such  scientific-computing workloads exhibit   unique characteristics---high entropy in data, high-dimensional data, and tightly-coupled computing patterns---that render  na\"ive in-network approaches ineffective. 
 Our  key insight here is that due to these characteristics,   canonical approaches (e.g., for caching and indexing)  fail because they are largely ``open loop'' and lack runtime feedback to in-network services. For example, due to  rapidly changing key distribution in scientific simulations, existing in-network load balancing (e.g.,~\cite{sosp17-netcache}) schemes do not work effectively as they do not tackle  such dynamic changes in workloads.

In this paper, we present   \sys{} (Network-accelerated ScIeNtific Computing), an architecture for fully unleashing the potential benefits  of  in-network computing for this scientific workload domain.     To address the aforementioned shortcomings, we argue for   a closed-loop control architecture for in-network acceleration services. At a high-level, \sys{} collects application-specific information in the data plane using telemetry, computes useful information in the control plane, and provides the information to services as a feedback via northbound APIs. Closed-loop feedback provided by \sys{} can make various  in-network acceleration services effective by allowing them to react dynamic and demanding scientific workload at runtime.

We highlight two fundamental challenges in realizing this vision: (1) the need for network-wide visibility over high-dimensional data and (2) coping with a tight timing constraint of scientific workloads. To tackle these challenges, we present our preliminary ideas on using sketch-based multidimensional telemetry and a hierarchical control plane that can flexible control the trade-off between freshness and coverage of telemetry data.  To demonstrate the potential benefits of \sys{}, we showcase three concrete use cases.



While our exploration is no doubt preliminary, we hope that by drawing attention to this relatively unexplored  class of workloads,  our work  opens  up a range of exciting new research opportunities and directions  for the  community. These  include the design of  efficient management and synchronization of telemetry data, coping with the heterogeneity of  network devices, and making it fault tolerant.



\section{Background and Opportunities}
\label{sec:background}

This section describes the characteristics of three canonical scientific computing applications (\S\ref{sec:canonapps}), and relevant opportunities in computational science that can be realized by leveraging in-network computing (\S\ref{sec:opportunities}).

%
%


\subsection{Examples of Scientific Applications}
\label{sec:canonapps}

\begin{figure}[t]
    \centering
    \includegraphics[trim={0cm 3cm 5cm 0cm},clip=true,width=0.97\linewidth]{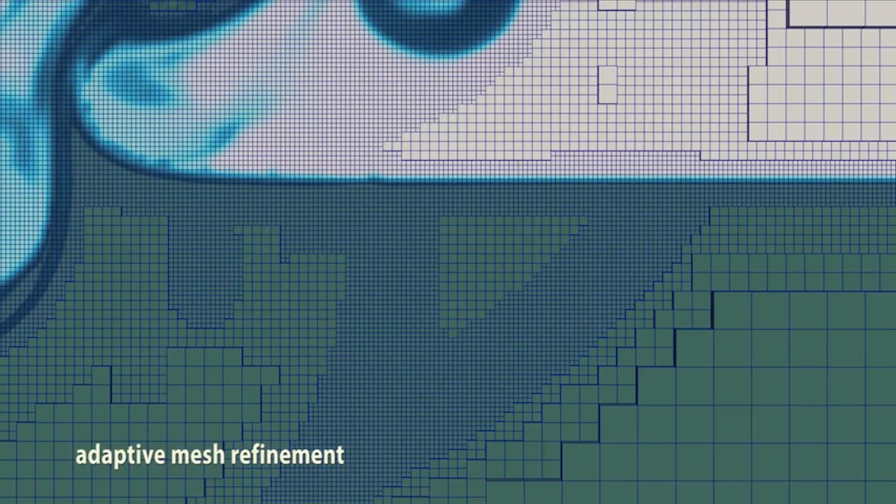}
    \caption{AMR mesh simulating an asteroid striking an ocean. Different colors of each cell indicate different materials (e.g., water or air). To accurately resolve the system state without high memory and processing overheads, xRage periodically subdivides high entropy cells (e.g., where the ocean is aerosolized) and merges low entropy cells.}
    \label{fig:amr-tsunami}
    \vspace{-15pt}
\end{figure}

\mypara{Understanding Turbulent Mixing}
xRAGE is a radiation-hydrodynamics simulation code for 
high energy-density phy\-sics \cite{Gittings_2008}. xRAGE and similar codes \cite{Fryxell_2000,van_der_Holst_2011}
rely on the 
adaptive mesh refinement (AMR) simulation technique. As shown in Figure~\ref{fig:amr-tsunami}, these codes decompose a physical region of space into a mesh of cells that are dynamically split and merged during the simulation to balance accuracy with processing overhead. AMR is critical at scale, where petabytes of memory would become a necessity without it.
Simulating the evolving mesh state requires frequent communication and synchronization between processes because neighboring cell attributes
(e.g., energies, pressures) are required to calculate each cell's state. When the entropy within a cell grows too large, that cell is split into a set of smaller cells and when the entropy within neighboring cells is both low and similar, those cells are merged. These merge/split operations are critical to making timely simulation progress by mapping similar amounts of work onto processors and network links, but it requires additional communication beyond that required to calculate the state of the simulation. Typical simulations require hundreds of network messages per process per second, and can scale up to hundreds of thousands of cores, thousands of network links, and run for months. 

\mypara{Simulating Magnetic Reconnection} 
VPIC \cite{vpic13} is an optimized particle simulation framework. In VPIC, a region of physical space is decomposed into a static mesh filled with moving particles.
In a VPIC simulation, each process represents a bounding box in the physical simulation space that particles move through. 
Large-scale VPIC simulations have been conducted with trillions of particles generating petabytes of data \cite{vpic13, vpic15, Zheng18} to support 
scientists exploring phenomena related to the  
magnetic reconnection that occurs during plasma formation~\cite{daughton10}.
Communication patterns within VPIC are centered around advancing particles and calculating diagnostics. At every timestep, each process communicates particle vectors to neighboring processes, effectively advancing the position of the particles through time. Because particles are represented with a small set of floating point numbers (32--64 Bytes of data), batch-based communications are critical to scaling VPIC. Diagnostic communications are customized explicitly for the evaluated hypothesis and are typically performed as \emph{collective communications} that require multiple subsets of processes to participate in synchronous reduction, scatter/gather, or similar network algorithms. For example, a diagnostic to calculate electric flux at varying scales requires all-to-all collective exchanges and collective all-reduce communications.



\mypara{Mapping the Genome}
Determining the specific order of DNA molecules in a genome, called \textit{sequencing}, is a fundamental step for biological applications such as personalized medicine~\cite{hipmer}. The most widely used instruments for DNA sequencing read only short fragments
randomly sampled from an input genome.
Genome assembly is the computational process of piecing together partially overlapping fragments to infer the most likely sequence of the full-length genome, and is one of the most demanding challenges in bioinformatics. 
The computational cost of genome assembly is dominated by small memory transactions within and between computing nodes. 
The HipMer assembly pipeline~\cite{www-hipmer, www-meraculous} begins with an input data set of short DNA reads randomly sampled from a genome~\cite{merbench}. Longer fragments of the genome are assembled by leveraging a domain-specific set of distributed hash tables to identify adjoining DNA sequences. Hash table operations are small in size, and during the assembly process some operations have strict order dependencies. HipMer’s communication patterns to maintain the hash table contents include: all-to-all exchanges, irregular lookups, and atomic updates. These patterns govern the efficiency of the parallel genome assembly pipeline at large scales where most of the assembly is communication bound.

\subsection{Opportunities for In-network Computing}
\label{sec:opportunities}

Scientific applications require frequent data movement and synchronization between nodes to evolve the large distributed state through time and to perform analysis on large scientific data sets. We observe that if some in-network capabilities could provide workload information collected from inside the network, it opens up multiple opportunities to dramatically improve the performance of the applications. In this section, we introduce three such opportunities.

\mypara{Accelerating Load-balancing and Mesh Adaptivity}
Unstructured meshes change dramatically over time with the computational and networking costs increasing as the number of cells grow, and with load migrating between processes as 
shocks and turbulence propagate. Both mesh refinement and load-balancing require additional communication with neighboring processes to describe changes to the mesh. Current approaches either pause the simulation to collect an accurate assessment of the distributed load and states of neighboring cells \cite{952105}, or use sampling to exchange less data \cite{10.1145/2287076.2287103,6877448}. An in-network telemetry approach to informing mesh adaptivity and load-balancing can assist in re-balancing the number of cells managed by each process, and provide more global data to enable merge/split cell operations without requiring more communication than sampling approaches.

\mypara{Accelerating Multidimensional Queries}
Queries on multi-dimensional data are important for identifying and tracking high-energy particles (using energy, position, and momentum) or detecting magnetic islands (using charge, position, and relative concentration). 
However, approaches optimizing query performance today can inflate I/O load by up to $42\times$ \cite{pebblesdb}, significantly affecting time to scientific discovery. Data indexes greatly improve query performance but require partitioned sets of data to pass through a single processing point so that compact indexed representations of the data can be constructed. 
First, leveraging network-based samples to construct distributions for attributes will enable applications to create efficient partitioning functions for scientific data without requiring an additional pass over the data set. These distributions are valuable in subsequent queries because only relevant partitions are included in the query scan phase. Second, in-network processing can be used to create summaries of the simulation communication to detect relative state information that is hard to determine using traditional techniques (i.e. identifying mesh regions that show large populations of identically charged particles exiting that region relative to nearby areas of the mesh).
The in-network construction of these distributed state approximations enables more efficient
detection and tracking of dynamic structures such as magnetic islands.

\mypara{Accelerating Data Matching and Scientific Analytics}
Scientific data analysis tasks, including aligning and matching fragments of DNA during genome sequencing can benefit greatly from an in-network caching service. However, typical caching lookup semantics do not apply to genome sequencing. A caching service for scientific analytics that includes the ability to provide and dynamically customize domain dependent rules for performing data matching will accelerate many classes of scientific analysis.
A domain-aware distributed cache routing layer that both routes a single lookup request to multiple candidate caches based on DNA alignment parameters (e.g., Smith-Waterman algorithm \cite{SMITH1981195}) and returns multiple candidate sequence locations offers the opportunity to make genome sequencing more efficient and simpler to implement.
Moreover, augmenting in-network services to provide telemetry data that tracks recently successful alignment and matching 
can further simplify and speed up genome sequencing.
\section{ Vision}
\label{sec:vision}
In this section, we highlight why off-the-shelf in-network acceleration techniques fail for scientific workloads. At a high level, we find these workloads have properties that make it hard to directly adopt existing solutions designed for conventional data center workloads. Based on this insight, we make a case for a ``closed-loop'' architecture to effectively realize these potential benefits highlighted in \S\ref{sec:background}.

\subsection{A Case for Closed-loop Control}
\label{sec:case}


The scientific applications introduced in~\S\ref{sec:canonapps} generate a massive amount of data modeled as key-value pairs, with up to hundreds of thousands of processes~\cite{byna2012}. For example, in scientific simulations such as xRAGE and VPIC, particles can be modeled as key-value pairs, with the pair representing a given particle and an attribute (e.g., energy, temperature).
While recent work has studied a wide range of in-network acceleration techniques for key-value store workloads in the context of data centers~\cite{sosp17-netcache, distcache-fast19}, we observe that existing approaches are not designed for the unique characteristics of scientific workloads. 
Here, we describe three such characteristics: entropy, high data dimensions, and tightly-coupled computing.

\begin{figure}[t!]
 \centering
 \includegraphics[trim={1cm 1cm 2cm 3cm},width=0.9\columnwidth]{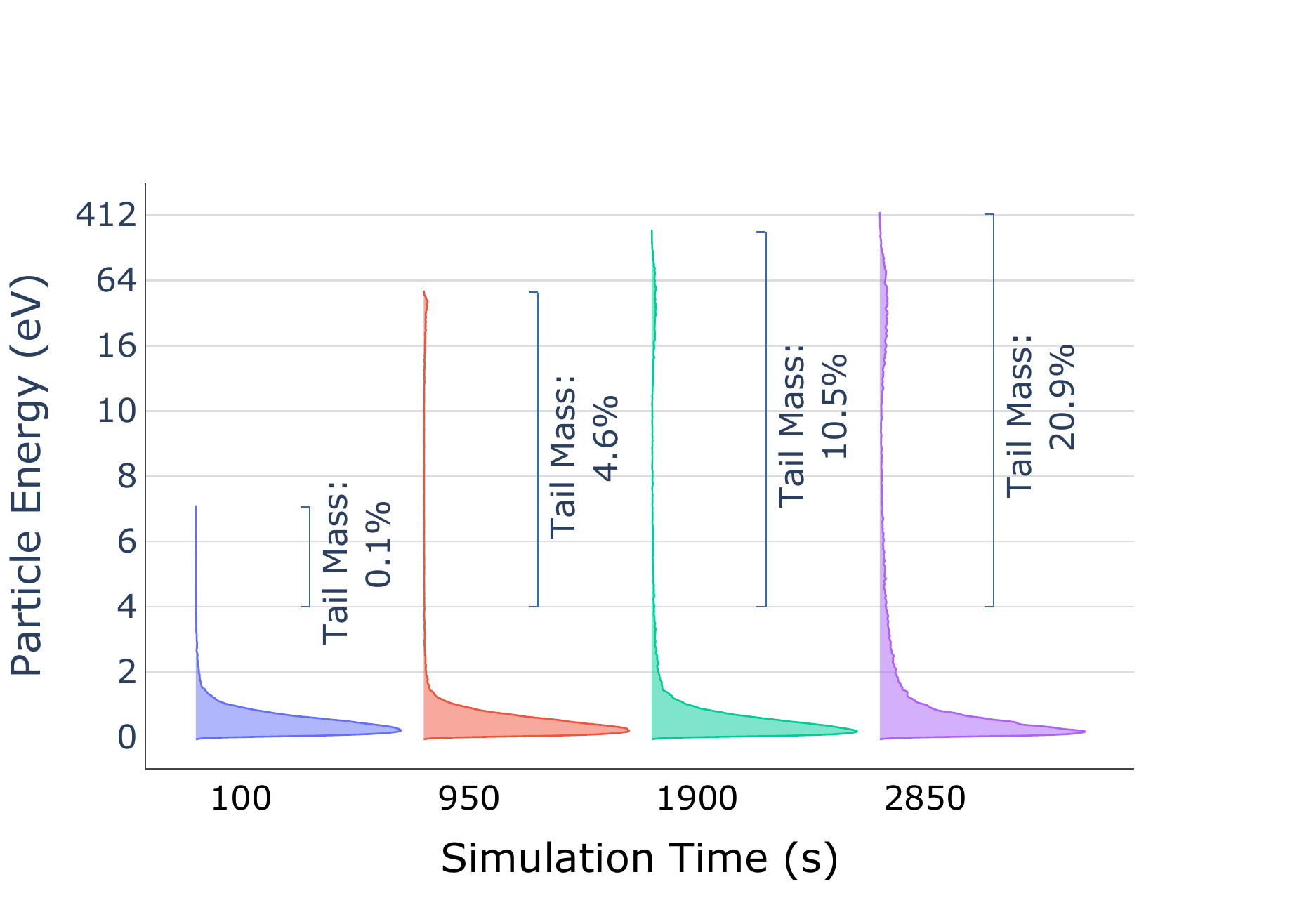}
 \vspace{-2pt}
 \caption{Distribution of particle energy during a VPIC Magnetic Reconnection simulation. Shown are the probability densities over time. The distribution changes rapidly, as energy is concentrated in a few particles.}
 \label{fig:nasi-distrib}
 \vspace{-10pt}
\end{figure}

\mypara{(1) High entropy} 
The distribution of keys in scientific simulations can change rapidly over time. In practice, this means that choosing any static partitioning of the key space is going to result in load imbalances (i.e., runtime slowdown) during the simulation. In Figure \ref{fig:nasi-distrib}, we show the probability densities for particle energy throughout a VPIC simulation. The rapid changes shown make it challenging to find a data layout that is efficient during both data generation and data querying time. It also increases the difficulty of finding caching algorithms that will maintain a high hit rate. 

\mypara{(2) High-dimensional data} Scientists attempt to record detailed observations of the characteristics of physical phenomena, so that detailed analyses can be performed even for aspects of a simulation that were not the initial target. As a result, scientific data is typically characterized by a large number of dimensions, e.g., temperature, energy, etc.
High dimensionality makes it challenging to index data so that queries to any of those attributes are answered fast.

\mypara{(3) Tightly-coupled computing} 
Many scientific applications consist of tightly coupled parallel processes that iterate together and require collective communication to make forward progress. The failure of one node to participate in this communication usually leads to a failure of the entire computation, and delayed responses can cause significant slowdown in the overall runtime of the application. As a result, the network requirements for tightly coupled applications are demanding and low latency operations are a top priority. 


Our analysis suggests that for many scientific workloads, due to their unique characteristics, existing in-network acceleration techniques may not work effectively. We observe that the root cause of the ineffectiveness is the lack of runtime feedback. For example, if we could provide the current load distribution observed inside the network as feedback to the system at runtime, it can be used to determine how to properly redistribute loads without significant communication overhead. The resulting timely reaction to load imbalances can help improve the overall performance while not affecting on-going network and compute workloads generated by high data dimensionality and tightly-coupled computing. This motivates us to envision a closed-loop control architecture for in-network acceleration for scientific workloads.


\subsection{Approach and Challenges}
We propose \sys{}, a closed-loop control architecture for in-network acceleration of scientific computing workloads.
At a high-level, \sys{} deploys telemetry in the data plane and  computes application-specific information (e.g., particle size distribution, entropy of moving data) in the control plane.
With this telemetry-based capability, \sys{} gives a closed-loop control to the acceleration services so that they can provide feedback to the system at runtime. Essentially, it provides a key building block for various in-network acceleration services that have the potential to significantly benefit scientific computing.
%
%

%
While  closed-loop control is not a new concept for networked  systems  (e.g., ~\cite{hotsdn14-bigdata}), we identify  key challenges in our specific context of using  in-network acceleration for scientific workloads:


\mypara{Challenge 1: Obtaining system-wide visibility on high dimensional data}
As mentioned earlier, scientific data is typically multidimensional. 
Recording and reporting full statistics for such multidimensional data in the data plane would require non-negligible amount of memory or could be even infeasible. Sampling can be an alternative, but it only provides a coarse-grained information with low accuracy. 
Also, since they focus on a single data plane device's view (e.g., switch) and ignore the network-wide visibility and aggregation, it is non-trivial to obtain network-wide visibility.

\mypara{Challenge 2: Coping with a tight timing constraint}
Even if we could get network-wide information, actuating acceleration services with the telemetry information in a timely manner is challenging. This is because most of scientific workloads typically require tighter timing requirements for the feedback loop.  For example, load balancing decisions have to be changed based on rapid key distribution changes which happens in a granularity of a few milliseconds. 


\section{\sys{} Design}
\label{sec:telemetry}

\begin{figure}[t!]
   \centering
   \includegraphics[width=\columnwidth]{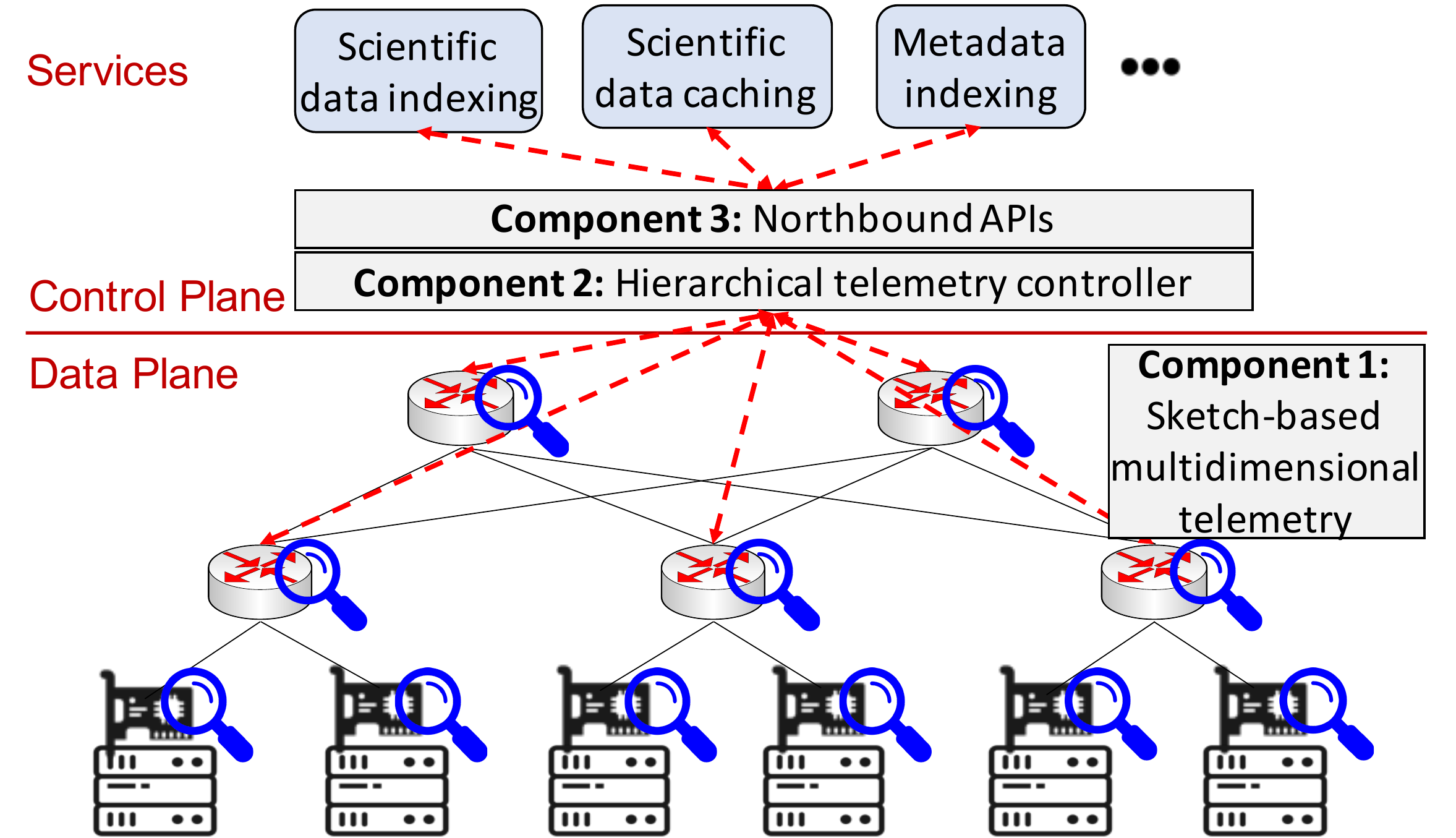}
   \caption{Overview of \sys{}.}
   \label{fig:overview}
   \vspace{-15pt}
\end{figure}
As a first step to realizing our vision and addressing the challenges, in this section, we describe our preliminary design of \sys{} as illustrated in Figure~\ref{fig:overview}. It consists of a sketch-based multidimensional telemetry in the data plane (\S\ref{sec:data_plane}), a hierarchical control plane    (\S\ref{sec:control_plane}), and northbound APIs (\S\ref{sec:api}).

\subsection{Sketch-based Multidimensional Telemetry}
\label{sec:data_plane}

To enable telemetry over high-dimensional scientific data efficiently, we argue for a sketch-based telemetry mechanism. Our two key insights lead to this direction: First, as prior work has demonstrated~\cite{univmon, OpenSketch}, sketching algorithms and data structures yield provable resource-accuracy trade-offs while enabling to obtain specific metrics of interest. Second, sketching algorithms are light-weight online algorithms and we observe that for scientific workloads, optimizing them with timely but  approximate information is better than with slower but  perfectly  accurate information because of the high dynamism in the workloads.

Based on these insights, we propose to leverage sketching data structures and algorithms for multidimensional telemetry in programmable data planes. Given the need to collect multiple application-specific statistics, we envision a sketching approach that can accurately estimate a set of metrics of interest at a low cost.
While there has been prior work on universal sketching such as UnivMon~\cite{univmon}, the high dimen\-sionality in scientific workloads brings additional challenges as universal sketching is designed to measure unidimen\-sional data. To begin with, we assign one universal sketch per dimension to track a broad spectrum of metrics (e.g., cardinality, change detection, and entropy) for that dimension. However, this strawman construction requires excessive amount of memory/compute resources as they are linearly increased in the number of dimensions. 

To reduce the memory/compute consumption, we propose:\\
\noindent{\em (1) Merged sketch structure.} Instead of maintaining separate equal-sized sketch structure for each dimension, we use a larger combined structure to maintain the counters for all dimensions. Since the workload distributions of all dimensions are not likely the same (low-error dimension can help high-error dimension), combined structure indeed increases the memory-accuracy tradeoff, similar to the analysis in~\cite{CountSketch}.\\
\noindent{\em (2) Merged compute.} For one dimension, universal sketches commonly update multiple counters by computing a series of hash functions, causing compute bottlenecks. As shown in prior work~\cite{RHHH,liu2019nitrosketch}, one can combine sampling techniques with sketching to reduce the number of hash computes and counter updates in software platforms. Their main idea is to uniformly sample a subset of counters to be updated. By repurposing this idea, when there are multiple dimensions to be updated, instead of drawing random samples independently for each dimension, we reuse the randomness across the dimensions and update to the same table of counters in the merged sketch. This is an equivalent process as the strawman in the algorithmic design but reduces the number of random number generations and hash function calls.


\subsection{Hierarchical Telemetry Controller}
\label{sec:control_plane}
Once each data plane device records sketches, we  need to periodically aggregate and analyze the telemetry data to inform future control actions. 
 A natural idea here is to use a centralized  controller that   computes application-specific metrics of interest with a network-wide view of the system, and provides  feedback for different scientific applications. While this centralized controller   provides a global view, it cannot provide a feedback in a timely manner, since the entire process in the feedback loop, including telemetry and computation, can take up to a few seconds.

To address this issue, instead of relying only on the central controller, we choose to deploy the controller in each switch's control plane (called a local controller) as well as the central one in a hierarchical manner and use the local controller to provide fast feedback to the services. This design is derived from our insight mentioned above that a timely less-accurate feedback (i.e., based on a local view) is more useful for scientific workloads. 

However, since each local controller only has access to locally collected sketches (e.g., within a rack), it may not be able to provide an useful feedback to services which require network-wide metric estimates. To resolve this problem, we let local controllers and the central controller synchronize the computed estimates periodically, making the controllers have an eventually consistent global view of the system. Each local controller periodically reports its own estimates to the central controller which aggregates reported local estimates and distributes the aggregate estimate back to local controllers. This way, each local controller eventually learns network-wide information as well as local information.       
 

\subsection{Northbound APIs for Services}
\label{sec:api}

\sys{} provides a set of APIs to expose the telemetry-based feedback capability to acceleration services. To get feedback from the telemetry controller using \sys{} APIs, an acceleration service needs to be deployed  both on top of the central and local controller. This way, the service logic on each switch can get more fresh but possibly local-oriented information while the one running on higher-level switches or central controller can obtain less-fresh but more network-wide information. Here, we introduce a few essential APIs and how services can use them:

\noindent{\em (1) set\_attributes (attribute\_list, estimate\_list, timing)}: It initializes a set of attributes and metrics of interest and a timing constraint (tight or loose). Based on these inputs, the controller deploys sketch structure and logic on each switch data plane, and a local controller on switch control planes if necessary (e.g., when the service requires a tight timing). This API is called by the service logic at the central controller.

\noindent{\em (2) get\_estimates (estimate\_buffer,  callback)}: The service uses this API call to retrieve a set of estimates from the controller to make a decision. It allocates a buffer where the estimates will be stored and  passes it as an argument. Once this API is called, when there are new estimates available, the callback function will be triggered so that the service can retrieve the estimates from the buffer.

Using the retrieved estimates, a service can make an appli\-cation-specific decision (e.g., resharding keyspace for indexing service). Based on the decision the service can update the system state (e.g., relocate key-value pairs). Also, the service can reconfigure attributes or estimates of interest via the set API call if necessary.

\section{Case Studies}
\label{sec:case_study}
In this section, we describe potential in-network acceleration services built on top of \sys{}, and we outline how they can directly benefit scientific applications.

\subsection{Network Accelerated Scientific Indexes}
\label{sec:research-index}

For scientists to derive insights efficiently, data queries must complete fast. Two common query types are point and range queries.
A \textit{point query} finds all particles matching a given key, whereas \textit{range queries} return all keys in a given range. Range queries are important for scientific applications, such as tracking
wavefronts in a shock tube.
Queries can be sped up by indexing data.
Recent work indexes data as it is written to storage \cite{Zheng18} but targets specifically point queries. Range queries today are supported through data structures that reorder data once written \cite{bayer72, oneil96, bender15}.
This can inflate I/O load by up to $42\times$ \cite{pebblesdb} and increase time to scientific discovery.

This problem can be mitigated by approaches that rearrange data for key locality \textit{before} it is persisted to storage, effectively trading off storage pressure for network bandwidth and in-network computation. This has the potential to reduce I/O load amplification significantly. 
%
%
A novel approach, however, would need to address the challenges we highlight in \S\ref{sec:case}, in order to support range queries on scientific data. Next, we describe how these challenges are relevant here. 

\mypara{Dynamic distribution detection} Existing approaches exclusively support point queries because load balancing and key locality preservation are hard to achieve together. Key hashing scatters data uniformly randomly across nodes, balancing load but destroying locality. As we show in Figure \ref{fig:nasi-distrib}, particle attribute distributions are not uniform, and can change rapidly over time due to the high entropy that is characteristic of scientific applications, so static locality-preserving partitioning approaches would fail to ensure balanced load as well. As a result, future work would have to dynamically re-shard the key space across nodes.

\mypara{Inter-node coordination} Each node’s visibility of the complete load distribution is limited. This is because every node only has access to keys directed to it, and due to the timing limitations of tightly coupled computing. Inter-node coordination is needed to observe the global key distribution.
Programmable network switches have visibility into ongoing communication, which provides a unique opportunity to offload at least a portion of the inter-node coordination process. Building the global key distribution in the network would partition load, while conserving resources at the compute nodes, and reducing network communication traffic. This coordination is heavily reliant on histogram building and merging, and should be a good fit for the in-network hardware capabilities, and thus maps well to \sys{}'s ability to enable distributed sketching techniques through its hierarchical architecture. The need to achieve this in a timely manner, and over multiple data dimensions, makes this a non-trivial enterprise.



\subsection{Network-based Scientific Data Caching}
\label{sec:research-cache}

Caching and memoization services are important acceleration techniques for scientific applications. Moreover, domain-specific caching services have broad applications in simulation and analysis codes that similarly rely on equation of state databases, fine-scale calculations, or other cached intermediate results. For example, HipMer's genome sequencing algorithm uses domain-specific distributed caches to assemble small, out-of-order and partially overlapping fragments of DNA into a single long genome. Adapting in-networking caching services to support scientific applications, however, introduces several new challenges not previously addressed in prior work on in-network caching services.

\mypara{Policy-driven cache control} In order to gain the greatest benefits from in-network caching services scientific applications must be able to modify in-network caches to use application-specific policies for eviction, replacement, and enable key lookup based on scientific parameters. Dynamically discovering cached data locations through some form of content-based routing would allow routing cache lookups based on proximity, and cache insertions could similarly be routed to locations with less network load. \sys{}'s design allows this to be achieved through the collection of relevant telemetry, and its aggregation at its telemetry controller.

\mypara{Distributed data structures} Discrete data structures like deques, sets, hash tables, and trees are essential in many scientific applications. These data structures fit naturally into shared-memory single-node architectures which limits scale. Distributing data structures across the memory of many nodes provides scaling, however data consistency and program semantics often limit the sequence and number of remote accesses in flight at any time. Scientific codes which frequently access distributed data structures to perform lookups or updates often result in inefficient network messaging. Latencies caused by small message exchanges decrease computational efficiency and increase time to solution, due to the nature of tightly-coupled computing that renders it vulnerable to stragglers.

\section{Discussion and Future Work}
We conclude by highlighting other  practical challenges and
  directions for future work.

\mypara{Dealing with heterogeneity of network devices}
We expect that scientific  clusters will have a mix of legacy non-program\-mable network devices that support simple sampling  tasks and modern programmable network hardware that can perform richer packet processing. Thus, a  practical solution must take into account deployment  heterogeneity  and yet provide accurate real-time telemetry capabilities. 


\mypara{Native multidimensional sketching}
Our preliminary design builds on existing  unidimensional sketch algorithms. However, for workloads whose dimensionality is too high, this may be inefficient. An open challenge is the design of  a native multidimensional sketching algorithms whose compute and memory requirements will be sublinear in terms of the number of dimensions.

\mypara{Fault tolerance}
Device failures can affect the accuracy of services built on top of in-network computing.Previous studies indeed have shown that such failures are prevalent in data centers~\cite{sosp17-crystalnet, imc18-fb_reliability}. To cope with such failures, we need to devise a way to make closed-loop control and telemetry  resilient to device failures.


\bibliographystyle{abbrv}
\bibliography{hotnets2020}

\end{document}